\documentclass[reprint,amsmath,amssymb,aps,pra,groupedaddress,showpacs]{revtex4-1}
\usepackage{latexsym,amscd,amsopn,amsthm,amsfonts}
\usepackage[dvips]{color,graphicx}%
\usepackage{dcolumn}%
\usepackage{bm}%

\def\e{\begin{equation}}
\def\f{\end{equation}}
\def\_#1{{\bf #1}}
\def\o{\omega}
\def\.{\cdot}

\def\E{\epsilon}

\def\r#1{(\ref{eq:#1})}
\def\=#1{\overline{\overline #1}}
\def\##1{{\bf#1\mit}}

\newcommand{\ka}{\hbox{\ae}}

\begin{document}

\title{Hyperbolic carbon nanoforest for phase matching of ordinary and backward electromagnetic waves: second harmonic generation }

\author{A. K. Popov}
\affiliation{Birck Nanotechnology Center, Purdue University,
West Lafayette, IN 47907,~USA}
\email{Corresponding author: popov@purdue.edu}
\author{I. S. Nefedov}
\affiliation{Aalto University, School of Electrical Engineering,
P.O. Box 13000, 00076 Aalto, Finland\\
Laboratory Nanooptomechanics, ITMO University, St. Petersburg, 197101, Russia}

\author{ S. A. Myslivets}
\affiliation{Kirensky Institute of Physics, Federal Research Center KSC SB RAS, 660036 Krasnoyarsk, Russia \\
Siberian Federal University, 79 Svobodny pr., 660041 Krasnoyarsk, Russia}


\date{}


\begin{abstract}%

We show that deliberately engineered dispersive metamaterial slab can enable the co-existence and phase matching of contra-propagating ordinary fundamental and extraordinary backward second harmonic surface electromagnetic modes. Energy flux and phase velocity are contra-directed in the backward waves which is the phenomenon that gives rise to unique nonlinear optical propagation processes. We show that frequencies, phase, and group velocities, as well as nanowaveguide losses inherent to the electromagnetic modes supported by such metamaterial,  can be tailored to maximize conversion of frequencies and to reverse propagation direction of the generated wave. Such a possibility, which is of paramount importance for nonlinear photonics, is proved with a numerical model of the hyperbolic metamaterial made of carbon nanotubes standing on the metal surface. Extraordinary properties of the backward-wave second harmonic generation in the reflection direction and of the corresponding frequency doubling metareflector in the THz are investigated with a focus on the pulsed regime.
\end{abstract}

\pacs{41.20.Jb, 42.25.Bs, 42.65.Ky, 42.65.Sf}
\maketitle
\section{Introduction}
Metamaterials (MMs) can enable exotic electromagnetic waves with contra-directed phase velocity and energy flux. Such waves are commonly referred to as backward electromagnetic waves (BEMWs). It was shown that BEMWs give rise to greatly enhanced optical parametric amplification and frequency up and down shifting nonlinear reflectivity \cite{APB}, to unusual features in the second harmonic generation (SHG) \cite{APB,KivSHG} and to extraordinary transient processes in the three-wave mixing of ordinary and BEMWs in a pulsed regime \cite{Tran}. Phase matching, i.e., equality of the phase velocities of the coupled waves is the requirement of a paramount importance to realize such advantages. Current mainstream in producing the BEMWs bases on the negative-index MMs (NIMs) \cite{ShC}. The negative-index resonance is usually narrow and phase matching of the ordinary and BEMWs presents a significant challenge \cite{Cai}. Quasi-phase-matching of contra-propagating ordinary waves is limited to the spatially periodically modulated nonlinear crystals. The possibility of the birefringent-like phase matching of ordinary EMWs in hyperbolic MMs was investigated recently in \cite{Dun}. An alternative paradigm for the realization of BEMWs can be implemented through the engineered negative dispersion $\partial\omega/\partial k<0$ \cite{AgGa,Agr}. Engineering the MM which can support opposite sign of the spatial dispersion at the fundamental and its SH frequencies and concurrently provide for their phase matching presents an even much greater challenge.  An approach to achieve the stated goal was proposed in \cite{APA2012}. It was based on the possibility to realize the negative dispersion $\omega(k)$ in the MM made of carbon nanotubes (CNTs) \cite{Tr,PNFA}.  This paper is to demonstrate the possibility of nanoengineering the MMs which support tailored electromagnetic modes that can satisfy to the whole set of the above-outlined requirements of the paramount importance: required matching frequencies, contra-directed energy fluxes, and equal co-directed phase velocities. Unusual properties of the corresponding frequency-doubling metareflector and advantages of SHG in such MMs will be also demonstrated through numerical simulations.

\section{Surface electromagnetic modes in the carbon nanoforest.}
\begin{figure}[h!]
\includegraphics[width=.6\columnwidth]{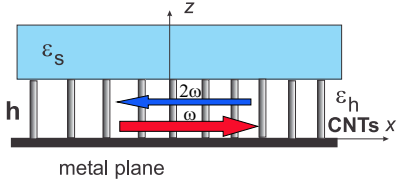}
\caption{Geometry of free-standing CNTs.} \label{geom}
\end{figure}

The underlying physical principle is based on the relationship ${\mathbf{S}}={\mathbf{v}_g}U$ between the Poynting vector ${\mathbf{S}}$, group velocity ${\mathbf {v}}_g={\rm grad}_{\bf k}\omega(\mathbf{k})$ and the energy density $U$. In the case of waves propagating along the axis x, it reads ${\mathbf{S}}=(\mathbf{S}/k)(\partial\omega/\partial k_x)$. It is seen that the the energy flux $\mathbf{{S}}$ becomes directed \emph{against} the wavevector $\mathbf{{k}}$ if the dispersion becomes negative: $\partial\omega/\partial k_x<0$. To realize such the opportunity, consider finite-thickness slab of metallic CNTs of height h embedded into a host matrix with relative permittivity $\E_h$ standing on a metal surface and open to the upper dielectric layer with relative permittivity $\E_s$  (Fig.~\ref{geom}). The nanotubes form a two-dimensional periodic structure in the $xy$-plane with the square lattice (for simplicity) and the lattice constant $d$.
Space-time dependence of fields and currents is taken as $\exp{[-i(\omega t-k_xx-k_zz)]}$.
For the simple case of  surface waves propagating in the given uniaxial slab of CNTs with open ends and the air as the host matrix, whose fields attenuate exponentially from the interface to the air ($\E_h=\E_s=1$), the dispersion
$\omega(k_x)$ is given by the equation \cite{PNFA}:
\e\label{eq:dis}
{k_z}\tan{(k_zh)}={\sqrt{k_{x}^2-k^2}},\, k_{x}^2=\E_{zz}(k^2-k_z^2),
\f
where $k$ is the wavenumber in the vacuum, $k_x^2>k^2$. Corresponding component of the relative permittivity $\E_{zz}$ is given by \cite{PNFA,PRB84}
\e\label{eq:p2}\begin{array}{cc}
 {\E_{zz}}=1-{k_p^2}/({k^2+i\xi k}), \, k_p^2={\mu_0}/{d^2L_{0}},
\end{array}
 \f
where, $k_p$ is the effective plasma wavenumber, $L_{0}$ is the effective inductance of the CNTs per unit length, the parameter $\xi=\sqrt{\E_0\mu_0}/\tau$ is responsible for losses, $\tau$ is the electron relaxation time.  Radius of the CNTs $r$ and the lattice constant $d$ are taken to be $r=0.82$\,nm and $d=15$\,nm. Then the indicated parameters are estimated as \cite{PNFA,PRB84}  $L_0=3.7\times10^{3}$ H/m, $k_p^2= 4.22\times 10^{10}$\,m$^{-2}$, $\tau$=3~ps. The effective plasma frequency  is calculated  as $\omega_p/2\pi$ = 58.7~THz. Thus the MM can be thought as a uniaxial free-electron plasma where electrons can move only along $z$-direction.
Apparently, it corresponds to a medium with hyperbolic dispersion at frequencies below the
plasma frequency $\omega_p=k_pc$, because $\E_{zz}<0$, whereas $E_{xx} =E_{yy}=\E_h=1$.

\begin{figure}[htb]
\centering
\includegraphics[width=.48\columnwidth]{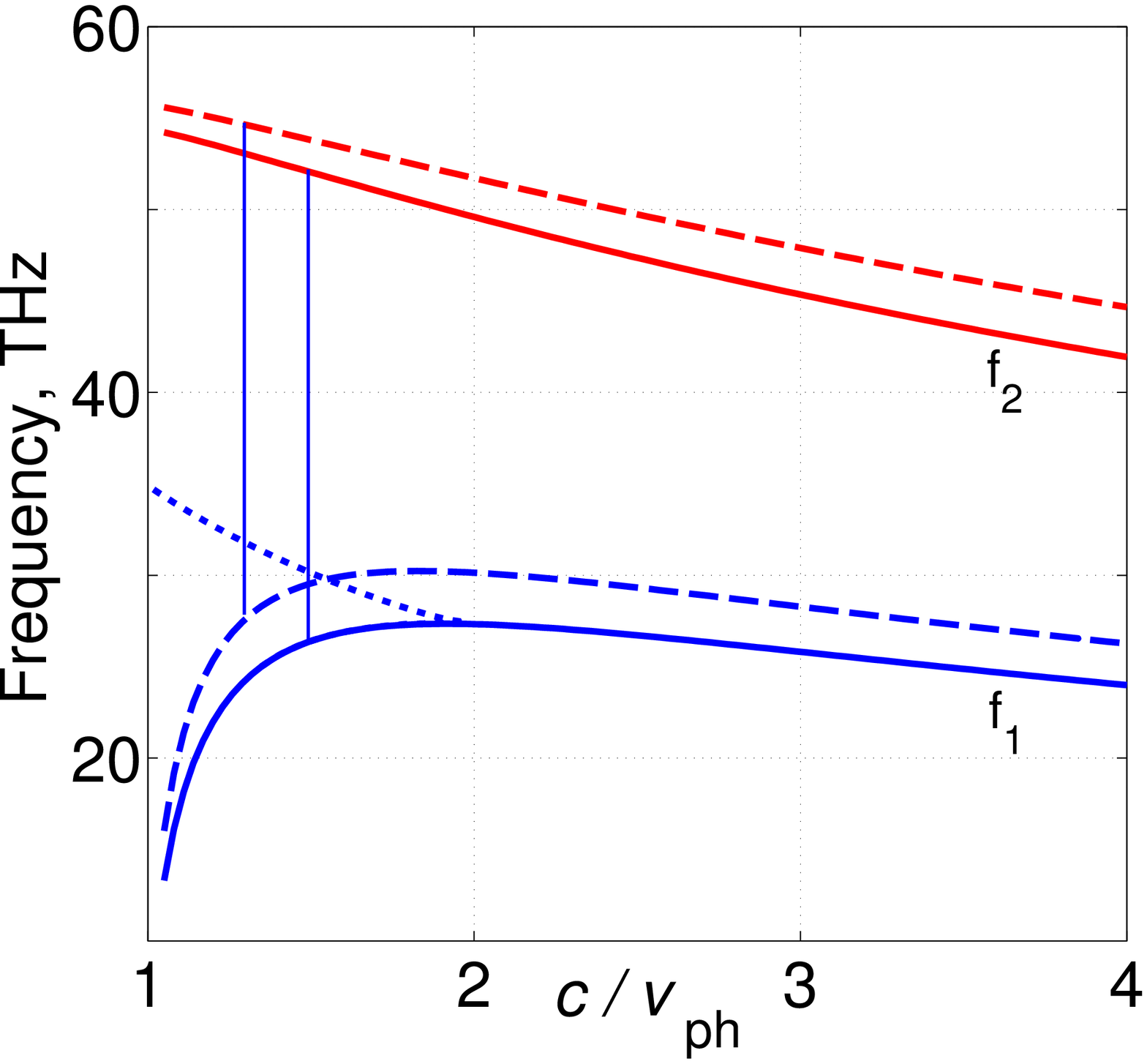}
\includegraphics[width=.5\columnwidth]{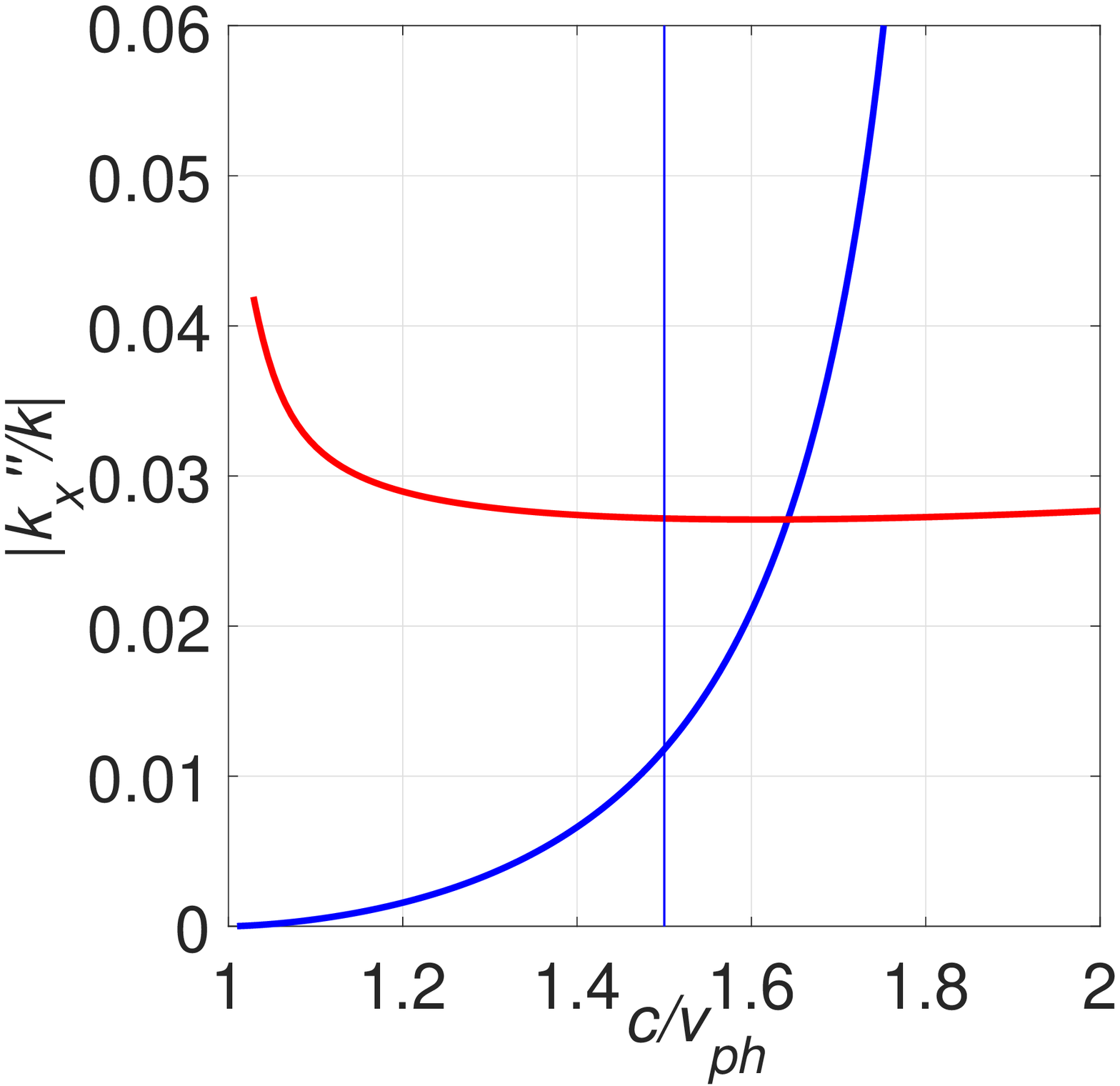}\\
(a)\hspace{0.5\columnwidth} (b)
\caption{ (a) Dispersion of two lowest eigenmodes in the slabs of standing CNTs with open ends. $\E_h=\E_s=1$; $h=1.05\, \mu$m (the solid lines) and $h=0.85\, \mu$m (the dashed lines).
 (b) Attenuation factor $k_x''$ for the lower-frequency mode (the ascending blue plot) and for the higher-frequency second mode (the descending red plot) at $h=1.05\, \mu$m.} \label{open}
\end{figure}

\begin{figure}[h!]
\includegraphics[width=.5\columnwidth]{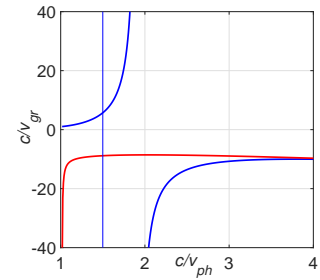}
\caption{Group velocity vs. phase velocity for the same two modes as in Fig.~\ref{open} for $h=1.05\, \mu$m. The the split blue plot is for the lower-frequency mode, the ascending red one is for the higher-frequency mode.}\label{group}
\end{figure}

We have calculated complex propagation constant $k_x=k_x'+ik_x''$  by numerically solving Eq.~\r{dis}.
Dispersion diagram for two lowest modes for two different thicknesses of the CNT layer is shown in Fig.~\ref{open}.
Here, reduced wave vector $k_x'/k=c/v_{ph}=n_{ph}$ is a slow-wave factor  which represents effective refraction index  $n_{ph}$. Real part of the normalized propagation constant $k_x'/k$ is shown for $h=1.05\, \mu$m (the solid lines), and $h=0.85\, \mu$m (the dashed lines). Vertical lines mark  frequencies of the phase matched ordinary fundamental and backward SH waves. The dotted line shows  real part of  the normalized propagation constant for complex wave, existing in the stop band and calculated for $h=1.05\, \mu$m.
It is seen that frequencies and dispersion of the allowed propagating EMW (eigenmodes) are determined by the
thickness and effective parameters of the metaslab and, hence, \emph{can be tailored}. Figure \ref{open}(a) demonstrates the possibility to adjust the eigenmodes so that the {ordinary} wave at frequency $f_1$ and the \emph{contra-propagating} {backward} wave at frequency $f_2$ travel with the \emph{equal} phase velocities. As stressed above, the later is the requirement of a paramount importance which gives rise to the \emph{extraordinary } SHG process.  A group delay factor $n_{\rm gr}=c/v_g$  for both modes is shown in Fig.~\ref{group}. It is seen that at $c/v_{\rm ph,1}=c/v_{\rm ph,2}\approx 1.5$, which corresponds to phase matching, $n_{\rm g,1}\approx 5.5$ and   $n_{\rm g,2}\approx 8.9$.

The dispersion displayed in Fig. \ref{open}(a) contrasts with that in the infinite uniaxial crystals \cite{Land}:
\e\label{eq:y1}
k_{\perp}^2=\E_{zz}(k^2-k_z^2),\, k_{\perp}^2=k_x^2+k_y^2,\f
which in the given case takes form
\e\label{eq:p8}
k_x^2=\left[1-{k_p^2}/({k^2+ik\xi})\right](k^2-k_{z}^2).\f
It also contrasts with the case of the MM made of CNTs standing between two metal planes:
\e\label{eq:w1}
 k_{x}^2=\E_{zz}\left[k^2-(m\pi/2h)^2\right].\f
Here, $k_z=m\pi/(2h)$, $h$ is the height of the waveguide, $m$ is a positive integer determining the number of field variations along CNTs. It is seen that in the latter case only BW propagation ($dk_x^2/dk^2<0$) is allowed at $\E_{zz}<0$ and $k<m\pi/2h$.
The appearance of the positive dispersion for small slow-wave factors and the stop-light regime $v_g\rightarrow 0$ in Figs. \ref{open}(a) and \ref{group} are explained by the fact that the Poynting vector in the upper bounding dielectric (here, air) is always co-directed with the wave vector, whereas inside the metaslab they can be contra-directed. The overall energy flow inside the metaslab is determined by the interplay of these fields which is dependent on the field variations across the metaslab and, consequently,  on the relation between $k_z$ and $k_x$.

Attenuation of EMWs  along the x-axis is represented by imaginary part of the propagation constant $k_x=k_x'+ik_x''$.  A magnitude of  $k_x''$ is  dependent on the electrons relaxation time in the CNTs $\tau$ at the frequencies below optical transitions.  Fig.~\ref{open}(b) shows attenuation of the both modes in the proximity of the frequencies that correspond to the phase matched fundamental and BWSH waves.  The branch of $f(c/v_{ph})$ which descends down to the stop light point on the dispersion curve (at approximately  27.5~THz for $h=1.05\, \mu$m)  is characterized by the complex constant corresponding to a huge damping. Its imaginary part,   $k_z''/k$, is not shown.  (Such a branch is not shown for $h=0.85\, \mu$m either and will not be considered further.)  It is seen that attenuation may become significantly different for different modes and frequencies and depends on the metaslab geometry.

 \section{Phase-matched backward-wave second harmonic generation in the carbone nanoforest. }
In the case of precise phase matching, normalized amplitudes $a_1$ of the first harmonic (FH) and of the SH, $a_2$,  are given by Maxwell's equations:
\begin{eqnarray}
 s_2({\partial a_2}/{\partial \xi})+({v_1}/{v_2}){\partial a_2}/{\partial \tau}=
  -igla_1^2-({\widetilde{\alpha}_2}/{2d})a_2,\quad \label{eq2a}\\
 s_1({\partial a_1}/{\partial \xi})+{\partial a_1}/{\partial \tau}=
     -i2g^*la_1^*a_2-({\widetilde{\alpha}_1}/{2d})a_1.\quad \label{eq2b}
\end{eqnarray}
Here,  $v_i>0$ and $\widetilde \alpha_{1,2}=a_{1,2}L$ are  group velocities and normalized attenuation indices at corresponding frequencies;   $\alpha_{1,2}$ are attenuation indices, $L$ is the metaslab thickness. Parameters $s_j$ take values $s_j=1$ for the ordinary wave, and $s_j=-1$ for the backward wave.
Quantities  $|a_j|^2$ are proportional to the time-dependent photon fluxes: $a_j=E_j/E_{10}$, where $E_j$ are slowly varying amplitudes of the electric components of electromagnetic fields,  $E_{10}$ is the amplitude of the FH field at the metaslab entrance. The approximation of plane traveling waves is employed. Coupling parameter $g=\ka E_{10}$, where $\ka=\sqrt{k_1k_2} 4\pi\chi^{(2)}_{2,\rm eff}$,
$\chi^{(2)}_{2,\rm eff}=\chi^{(2)}_{1,\rm eff}/2$ are effective nonlinear susceptibilities at the corresponding frequencies, $k_j=\omega_j/v_{ph}$, and $v_{ph}$ is the value corresponding to phase matching. We also define normalized metaslab thickness $d=L/l$,  position $\xi=x/l$ and  time instant $\tau=t/\Delta\tau$, where $l=v_1\Delta\tau$ is the pump pulse length, and $\Delta\tau$ is duration of the input FH pulse.

The shape of the input FH pulse  was chosen close to the rectangular profile:
\begin{equation}
F(\tau)=0.5\left(\tanh\frac{\tau_0+1-\tau}{\delta\tau}-\tanh\frac{\tau_0-\tau}{\delta\tau}\right).
\end{equation}
Here, $\delta\tau$ presents duration of the pulse front and tail, and $\tau_0$ is  a shift of the front relative to $t=0$.
The magnitudes $\delta\tau=0.01$ and $\tau_0=0.5$ were selected for numerical simulations.
The following values and estimates, which are relevant to the MM made of nanotubes of height  $h=1.05\,\mu$m (Figs.~\ref{open} and \ref{group}), were also used for numerical simulations. Spectrum bandwidth corresponding to the pulse of duration
$\Delta\tau=10$~ps is on the order of $\Delta f\approx 1/\Delta\tau=0.1$~THz. Hence,  $\Delta f/f \propto10^{-2}\div10^{-3}$, and  phase matching can be achieved for the whole frequency band. This becomes impossible at $\Delta\tau=10$~fs because of $\Delta f/f \propto10$ in this case.
Phase matching occurs at $k_1=5.47\times10^5$~m$^{-1}$, $k_2=2k_1$ (Fig. \ref{open}). Corresponding attenuation factors are calculated as $\alpha_1=2k_1^{''}=2(9.3\times10^{-3})k_1=1.02\times10^4$~m$^{-1}$, $\alpha_2=2k_2^{''}=2(2.72\times10^{-2})k_2=5.96\times10^4$~m$^{-1}$.
Since losses for the second mode is greater, the characteristic metaslab thickness corresponding to extinction $\exp(-\alpha_2L)=0.1$, i.e., to $\alpha_2L=2.4$, $\alpha_1L=0.41$,  is estimated as $L\approx40\, \mu$m. The FH pulse  length is estimated as $l=\Delta\tau v_1=\Delta\tau c/n_{g,1}=606\, \mu$m, which is  15 times greater than $L$. The later indicates that the quasistationary process establishes through almost the whole pulse duration, whereas some transients occur at the pulse forefront and tail. Note, that at $\Delta\tau\leq10$~ps, which is still acceptable, the effect of the transient processes significantly increases.

\begin{figure}[htbp]
\begin{center}
\includegraphics[width=.49\columnwidth]{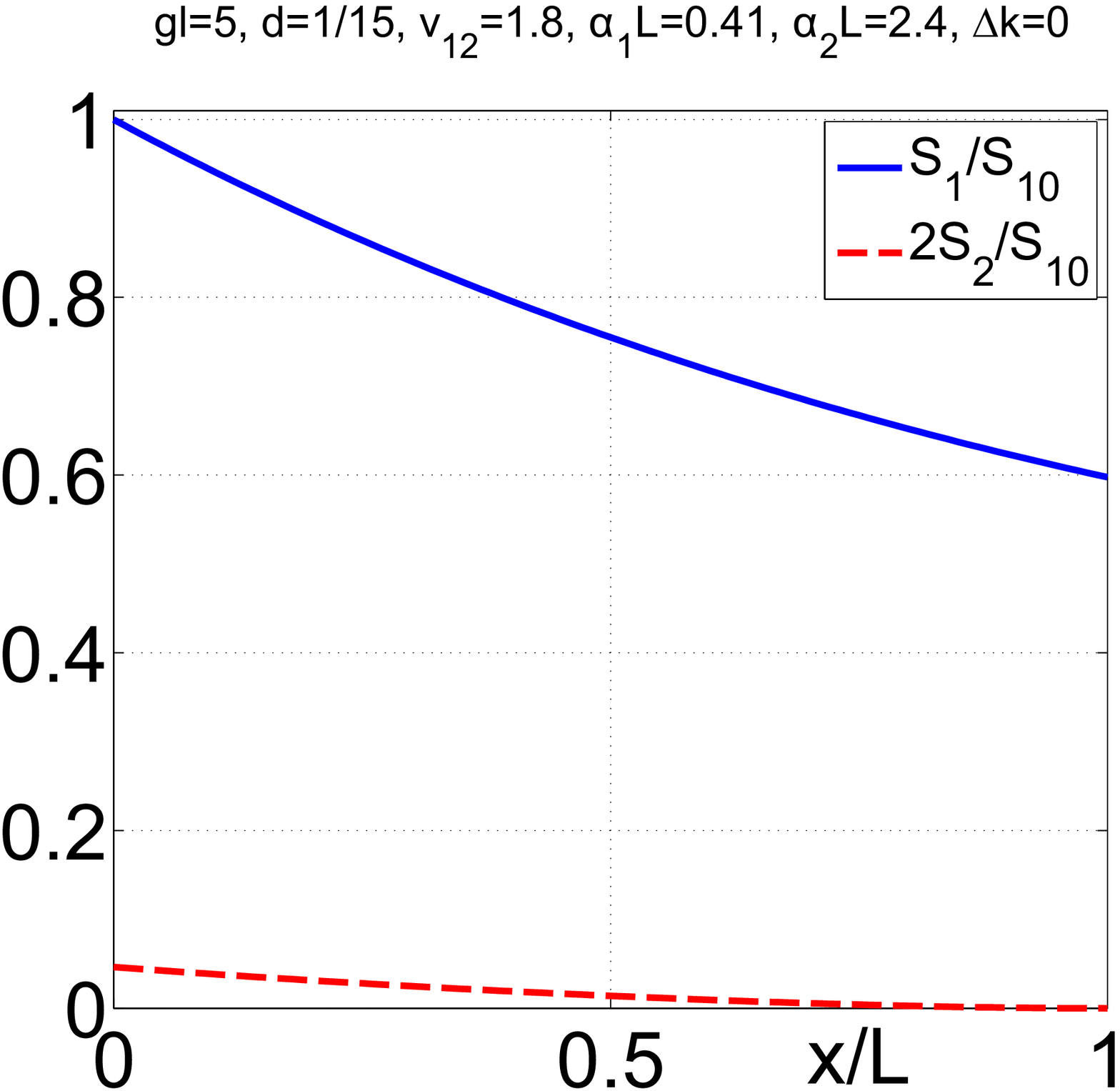}
\includegraphics[width=.49\columnwidth]{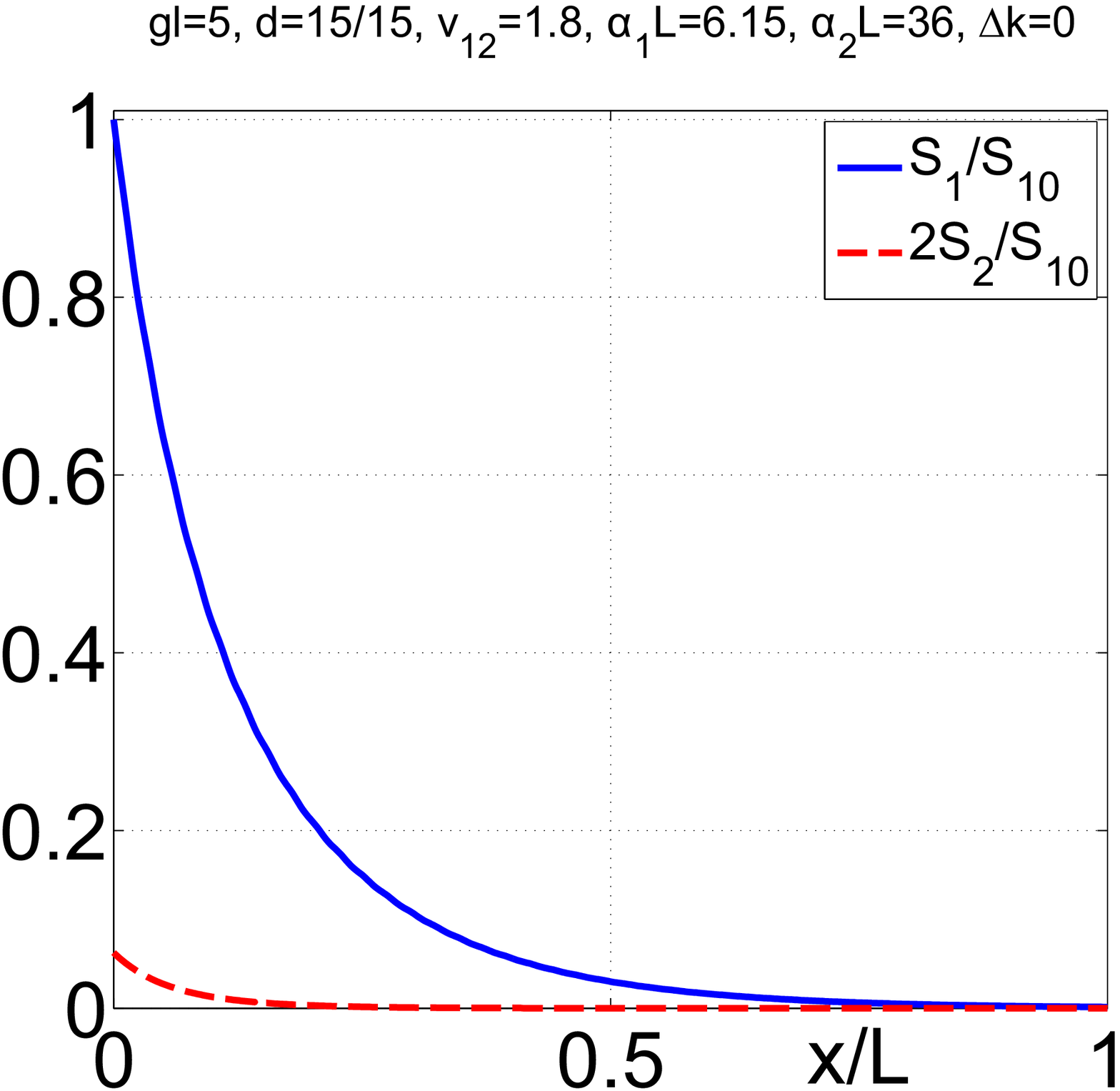}\\
 (a)\hspace{0.5\columnwidth} (b)\\
\includegraphics[width=.49\columnwidth]{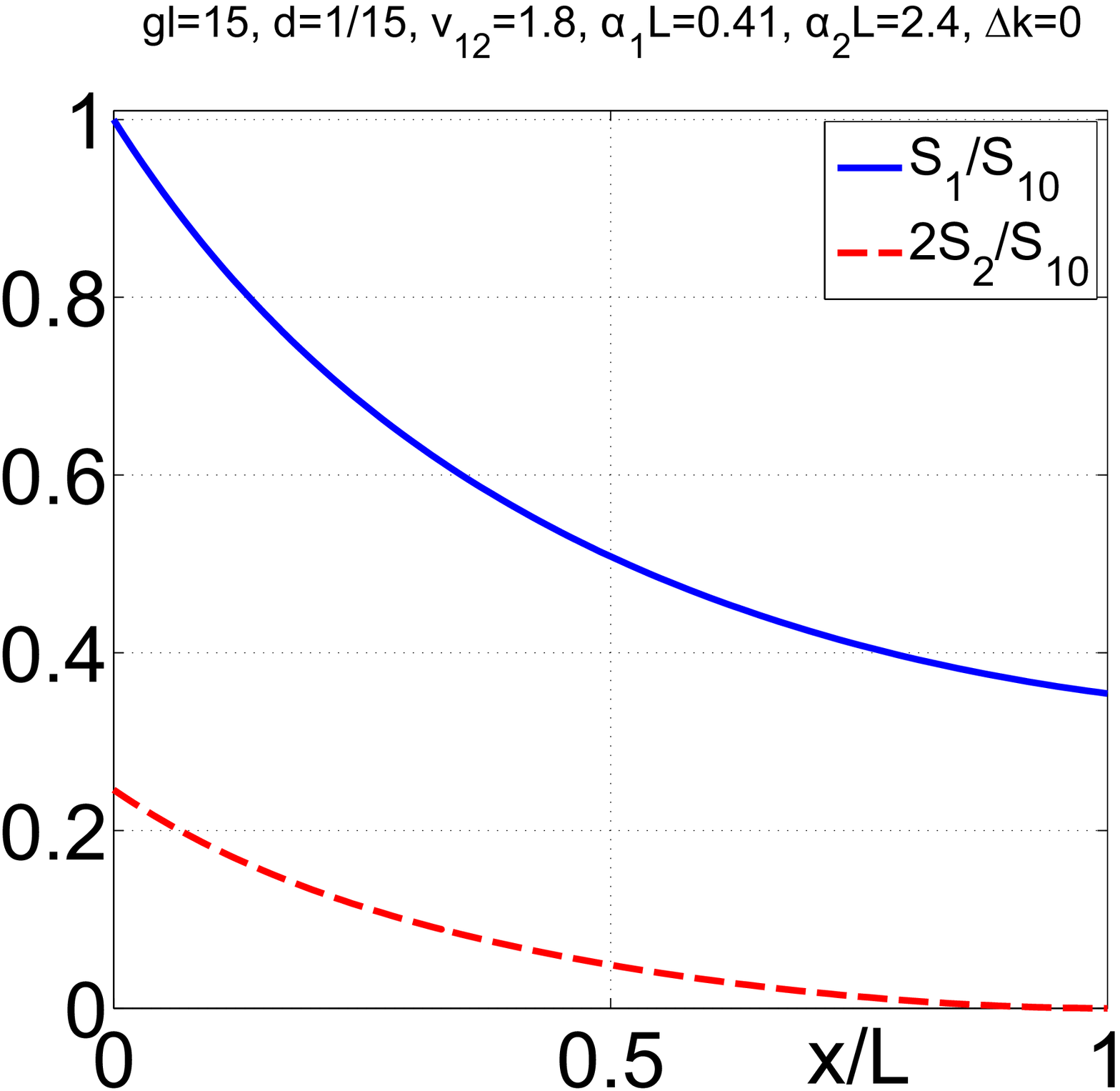}
\includegraphics[width=.49\columnwidth]{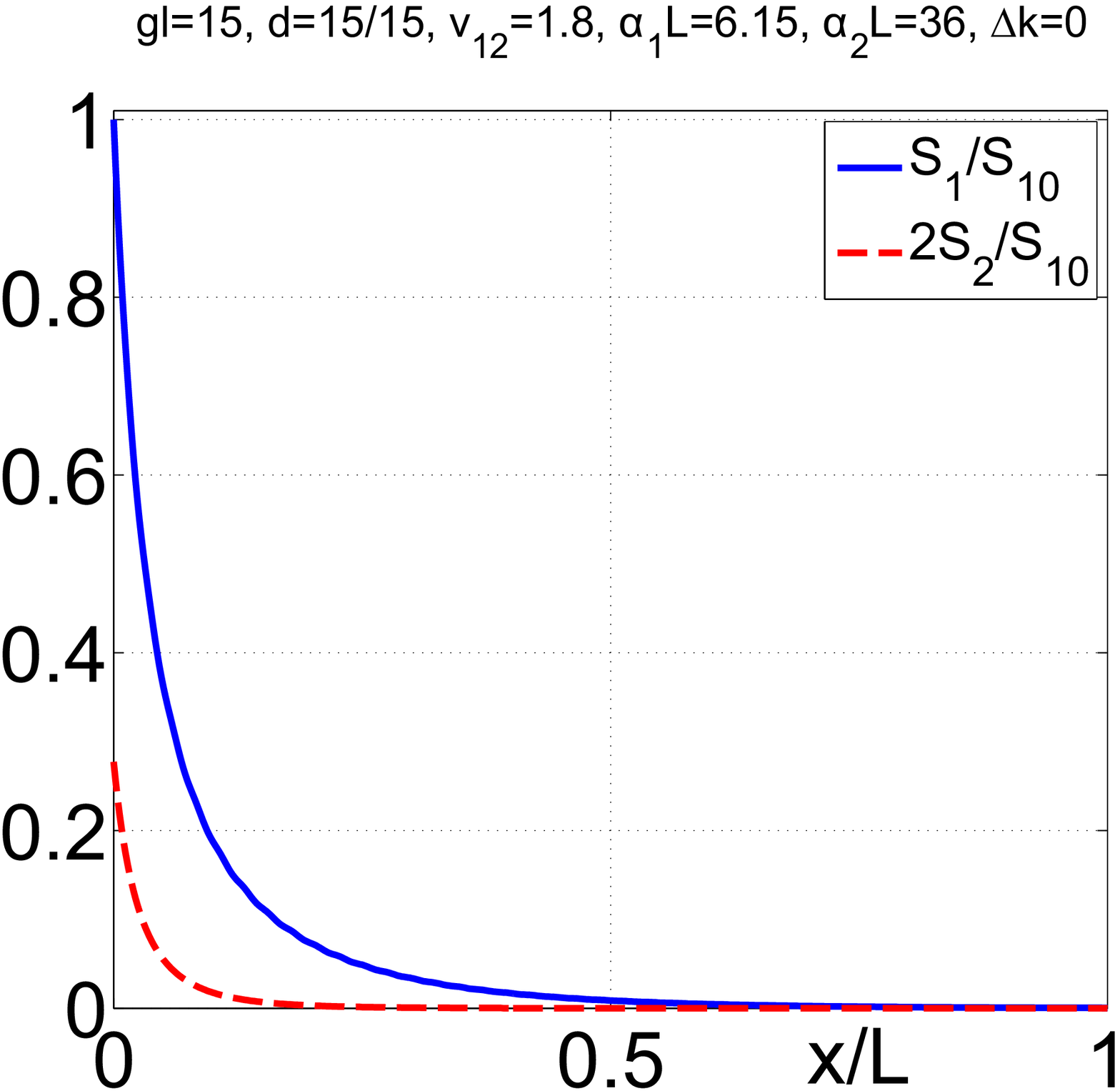} \\
(c)\hspace{0.5\columnwidth} (d)
\caption{BWSHG:  dependence of the energy conversion efficiency on  the metaslab thickness, intensity and duration of the pump pulse. (a) and (b): $gl=5$; (c) and (d): $gl=15$. (a) and (c): $L/l=1/15$; (b) and (d) $L/l=1$.} \label{nim}
\end{center}
\end{figure}
\begin{figure}[htbp]
\centering \includegraphics[width=.49\columnwidth]{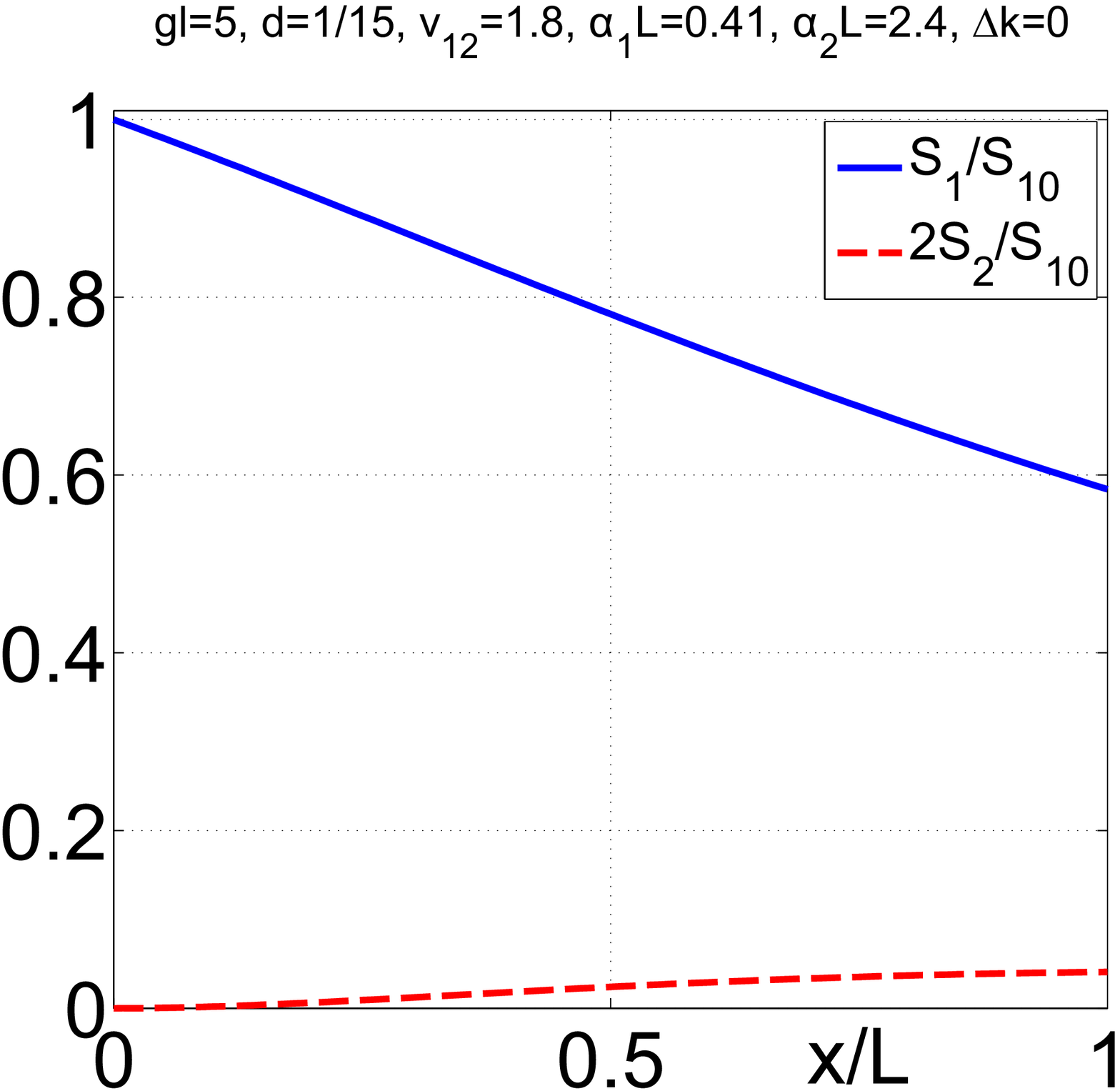}
\includegraphics[width=.49\columnwidth]{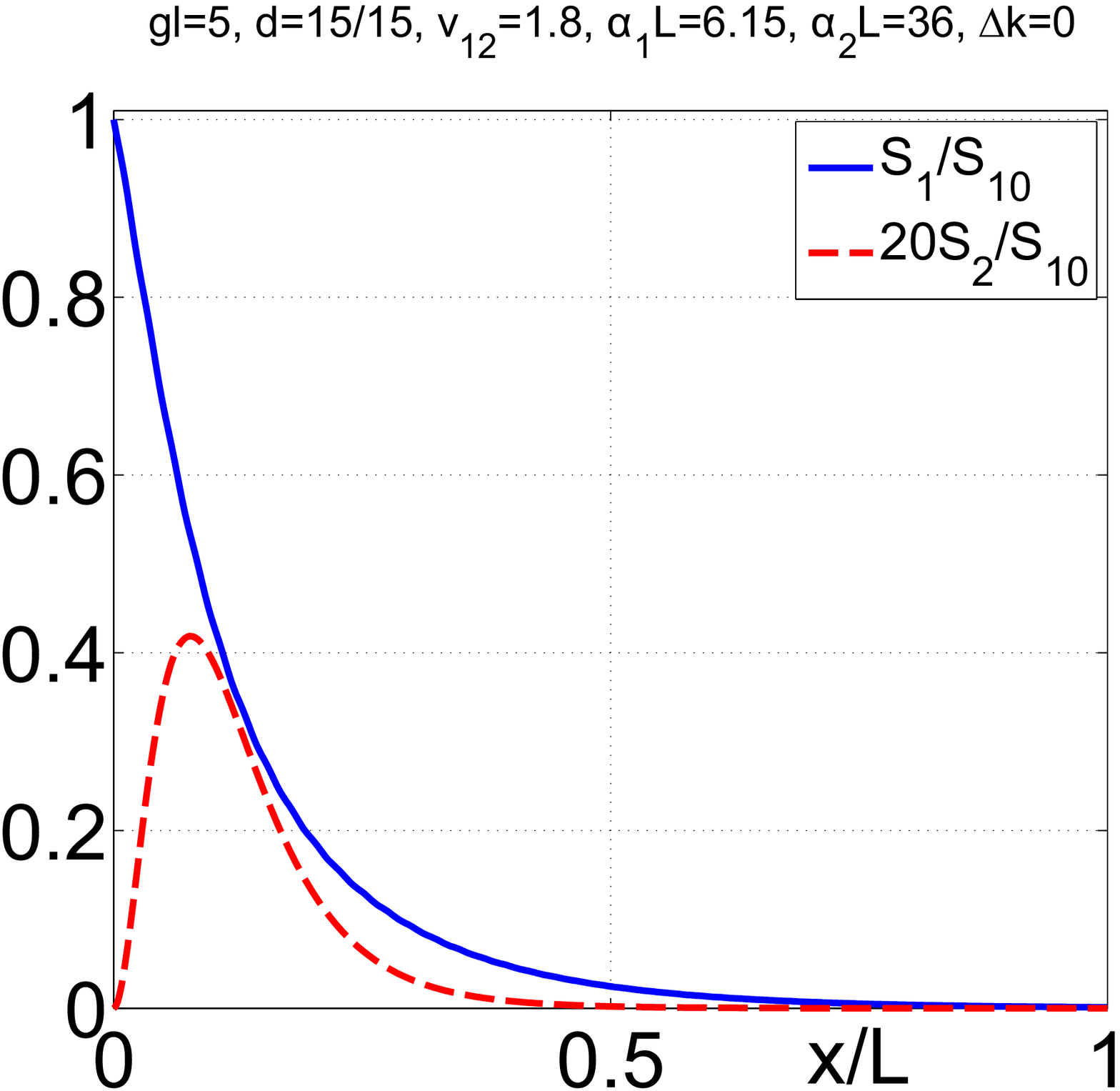}\\
\centering (a)\hspace{0.5\columnwidth}(b)\\
\centering
\includegraphics[width=.49\columnwidth]{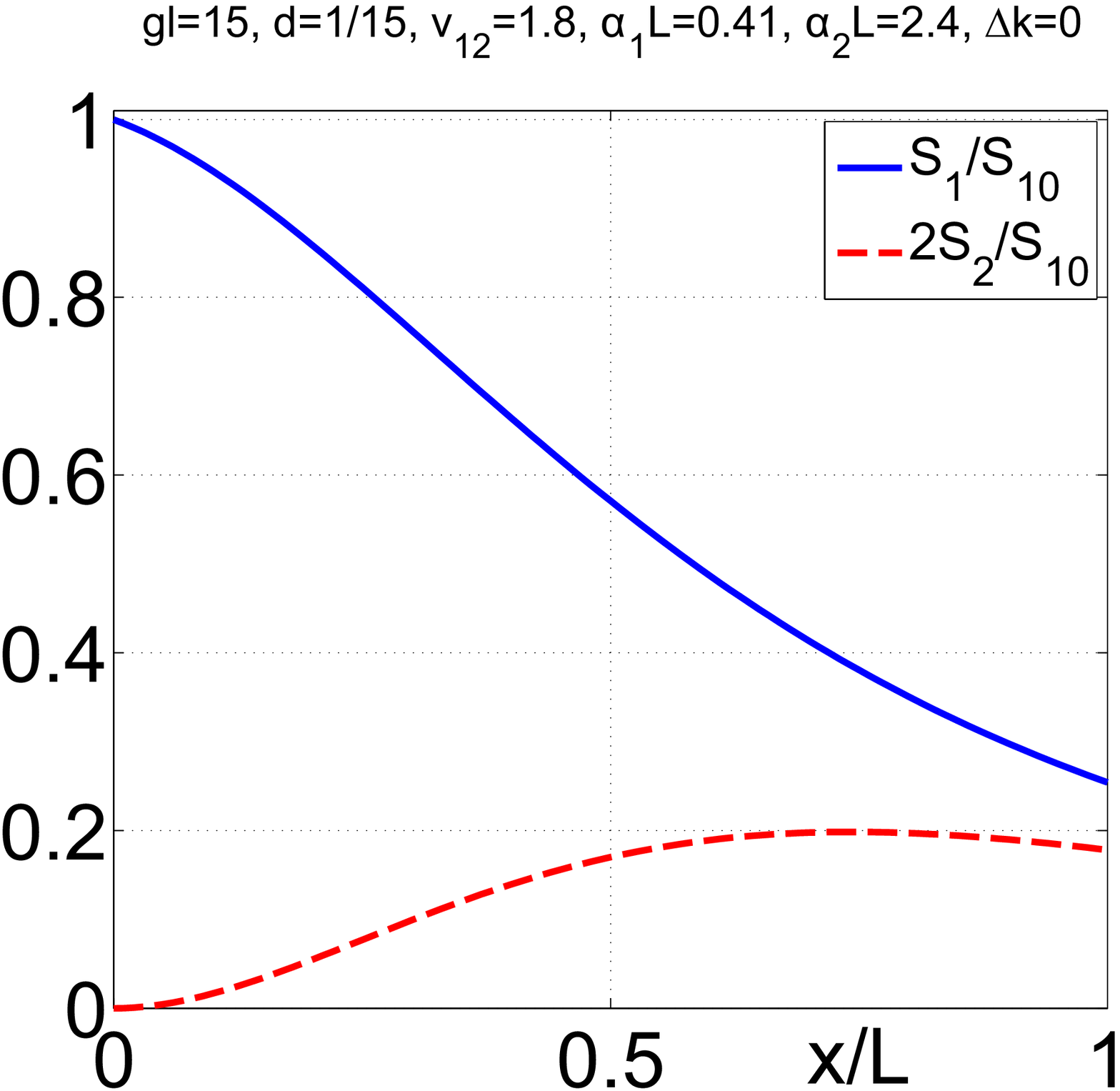}
\includegraphics[width=.49\columnwidth]{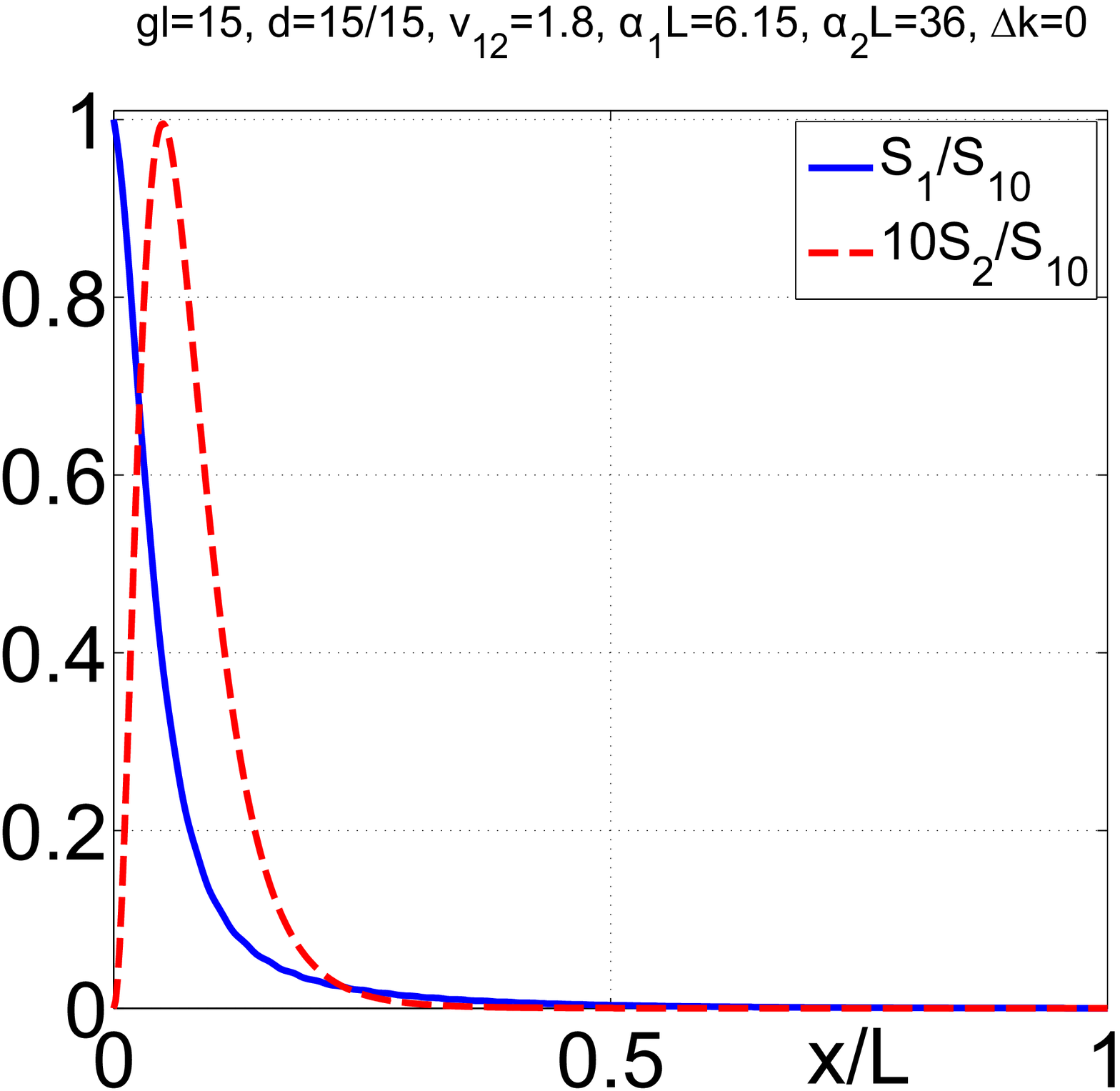}\\
\centering (c)\hspace{0.5\columnwidth} (d)
\caption{Ordinary SHG at all other parameters the same as in Fig. \ref{nim}.  } \label{pim}
\end{figure}

Figure~\ref{nim} presents the results of numerical simulations for energy conversion efficiency at BWSHG with an account for the above-calculated losses and group velocities. Here,   $\eta_2(x)=S_2(x)/S_{10}=\int dt |a_2(x,t)|^2/\int dt|a_{10}(t)|^2$  is the pulse energy (quantum) conversion efficiency and the factor $S_1(x)/S_{10}$ presents depletion of energy of the FH pulse  along the slab and at the corresponding exits: $x=0$ for the SH and $x=L$ for the FH. Two  coupling parameters  ($gl=5$ and $gl=15$) and two different input pulse lengths ($L/l=1/15$, and $L/l=1$) were chosen for the simulations. Coupling parameter $gl$ is proportional to the total number of photons per input FH pulse. It can be also thoughts as the ratio $l/x_0$ of the input pulse length $l$ and the characteristic slab thickness $x_0$ required for significant photon conversion from FH to SH for the given pulse intensity at its maximum. The interplay of several processes contributes to the outlined dependencies.  Figure~\ref{nim}  shows that conversion efficiency grows with increase of the input pulse amplitude. However, the important \emph{unusual} property is that  BWSHG, i.e., \emph{frequency-doubling nonlinear reflectivity} rapidly saturates with an increase of the metaslab thickness.   Such unusual behavior is due to the backwardness of SH which propagates against the FH beam and is predominantly generated in the area where both FH and SH are not yet significantly attenuated. It is seen that the overall nonlinear reflectivity provided by such a  frequency-doubling meta-reflector may reach the values on the order of ten percents at the selected values of the parameter $gl$.  Calculations also show that the reflectivity in the pulse maximum for the same parameters appears two times greater than the time-integrated values. These dependencies are in stark contrast with SHG in ordinary materials as seen from comparison with Fig.~\ref{pim}.
It  displays corresponding dependencies in the case of ordinary material with all other parameters the same as in Figs.~\ref{nim}. Here, both FH and SH exit the slab at $x=L$.  It is seen that, in general, SH reaches maximum inside the slab which is due to the interplay of nonlinear conversion and absorption processes. In order to to maximize the SH output, the pump strength, its pulse duration and the slab thickness must be carefully optimized in this case as appeared in the case of Fig.~\ref{pim} (c). Investigations show that shape and width of the output pulses in the cases of ordinary SHG and BWSHG also appear significantly different.

\section{Conclusions.}
We show the possibility to engineer the metamaterials that satisfy to a set of requirements of a paramount importance for the realization of extraordinary nonlinear photonic processes and devices which enable changing photons frequency and propagation direction. The proposed metamaterials support a set of travelling electromagnetic waves i) which frequencies satisfy to energy conservation law for nonlinear-optical frequency-conversion processes; ii) some of them are extraordinary backward waves with contra-directed energy flux and phase velocity, whereas other(s) are ordinary waves; iii) contra-propagating waves have equal phase velocities, i.e., are phase matched; iv) such properties can be adjusted to different frequencies. Frequency mixing of backward and ordinary waves possess fundamentally different properties compared to they ordinary counterparts and have important breakthrough applications in photonics. Current mainstream in crafting metamaterials that ensure backward waves relies on the engineering of mesoatoms -the nanoscopic LC circuits that provide a negative magnetic response at optical frequencies. The described in this paper approach is {fundamentally different} and bases on engineering the tailored coexisting negative and positive dispersion $\omega(k)$ of electromagnetic waves which dictates the particular relationship between the frequencies and wavevectors of normal electromagnetic modes.

Such a general possibility is demonstrated through numerical simulations making use a particular example of the "carbon nanoforest." It is the metamaterial made of carbon nanotubes of a particular diameter, height, and spacing standing on the metallic surface. We show that the negative and sign-changing dispersions pertinent to such metamaterial and  nanowaveguided surface electromagnetic waves can be tailored to support phase matched backward-wave second harmonic generation in the THz through near-IR frequency ranges. Attenuation introduced by the metallic properties of carbon nanotubes and by the particular nano-waveguide modes were investigated and appeared different for the coupled harmonics.  Most practically important, pulsed, regimes of second harmonic generation in such metamaterials were investigated with the simplified model of plane traveling waves. A set of coupled partial differential equations was employed which accounted for dispersion of group velocities and attenuations of the coupled pulses. Since the generated second harmonic travels in the direction opposite to the fundamental wave, the investigated process presents a model for the realization of the miniature frequency doubling metareflector/metaswitch and of the remotely controlled metasensor with unique properties. It was demonstrated by comparison with the ordinary second harmonic generation at otherwise similar conditions.

The described approach can be generalized to engineering the metamaterials of different architecture and composed of different materials which  support the tailored positive and negative dispersion of electromagnetic waves to enable extraordinary phase-matched coherent nonlinear optical propagation processes through an extended frequency band.

\textbf{Acknowledgements}
This material is based upon work supported in part by the U. S. Army Research Laboratory and the U. S. Army Research Office under grant number W911NF-14-1-0619 and by the Russian Foundation for Basic Research under grant RFBR 15-02-03959A.\\
\\

\appendix
\section{Derivation of the transfer matrix and dispersion equation}


 \subsection{$2\times 2$ transfer matrix for a slab of anisotropic material}

Let us consider anisotropic medium, characterized by the diagonal permittivity tensor (written in dyadic form), where the anisotropy axis is directed along the $z$-axis.
\e\label{eq:a1}
\=\E=\E_{\perp}(\_u_x\_u_x+\_u_y\_u_y)+\E_{zz}\_u_z\_u_z.
\f
Assuming that the  medium  is homogeneous along the $x$-axis and no field variation along the $y$-axis, we can write the time-space dependence as $\exp{i(-\o t+k_xx)}$ (along the $z$-axis the medium may be non-homogeneous).
Then $\partial/\partial\o=-i\o,\;\partial/\partial x=ik_x$, and $\partial/\partial y=0$.
Under such conditions Maxwell equation are split into two independent sub-systems, describing the TM-modes with $H_x=0,\,H_z=0,\,E_y=0$, and other nonzero components, and the TE-modes with $E_x=0,\,E_z=0,\,H_y=0$, and other nonzero components.
For each sub-system Maxwell equations can be written in form of a system of two ordinary differential equations for tangent field components. For the TM-modes:
\begin{eqnarray}\label{eq:a2}
{d}E_x(z)/{dz}&=&i\eta k\left[k-({k_x^2}/k)\E_{zz}\right]H_y,\nonumber\\
{d}H_y(z)/{dz}&=&i(k/{\eta})\E_{\perp}E_x
\end{eqnarray}
with $E_z=-({k_x}/{k}\E_{zz})\eta H_y$ (the relative permeability $\mu$ is assumed to be 1).
For the TE-modes:
\begin{eqnarray}\label{eq:a3}
{d}E_y(z)/{dz}&=&-i\eta kH_x,\nonumber\\
{d}H_x(z)/{dz}&=&({i}/{\eta})\left[({k_x^2}/{k})-k\E_{\perp}\right]E_y
\end{eqnarray}
with $H_z=({k_x}/{k\eta})E_y$.
Since in our geometry $\E_{zz}<0$ and $\E_{\perp}>0$, evidently the TM-modes only exhibit hyperbolic dispersion and are of our interest. By this reason the TM polarization only will be considered.

The system \r{a2} can be rewritten in matrix form:
\e\label{eq:a4}
{d{\rm X}(z)}/{dz}=i{\rm [A]}{\rm X}(z),\f
with the column-vector ${\rm X}=(E_x,H_y)_{\rm col}$ and elements of the matrix [A], defined by Eq.~\r{a2}.
The solution of the Cauchy problem for this matrix equation looks as
\e\label{eq:a5}
{\rm X}(z)=e^{i{\rm [A]}z}{\rm X}(0).
\f
Here [M]$(z)=e^{i{\rm [A]}z}$ is the transfer matrix, or the matrix propagator, which connects the tangent field components in planes $z=0$ and $z$. In general case the transfer matrix can be computed using spectral representation of the matrix function. If the matrix [A] has a simple structure, the matrix exponent is expressed as 
\e\label{eq:a6}
e^{i{\rm [A]}z}=\sum_{l=1}^s e^{i\lambda_lz}\sqcap_{k\neq l}
\frac{{\rm [A]}-\lambda_k{\rm [I]}}{\lambda_l-\lambda_k},\f
where $s$ id the number of different eigenvalues $\lambda_k$ of the matrix [A] and [I] is the unit matrix.

In our case $s=2$ and the eigenvalues are the following:
\e\label{eq:a7}
\lambda_{1,2}^2=k_z^2=k^2\E_{\perp}-k_x^2{\E_{\perp}}/{\E_{zz}}.\f
Actually, Eq.~\r{a7} describes dispersion of the TM-waves in unbounded uniaxial medium.
Using Eq.~\r{a6} one can obtain expression for the transfer matrix [M] in explicit form:
\begin{eqnarray}\label{eq:a8}
M_{11}(z)=\cos(k_zz),&\quad&
M_{12}(z)=iZ\sin(k_zz),\nonumber\\
M_{21}(z)={i}\sin(k_zz)/Z,&\quad&
M_{22}(z)=\cos(k_zz), 
\end{eqnarray}
 where $Z=E_x/H_y=\eta{k_z}/{k\E_{\perp}}$ is the transverse wave impedance for the TM-waves.

 \subsection{Dispersion equation}

Let us consider the structure, shown in Fig.~1, main text. The tangent field components can be recalculated from the ground plane $z=0$ to the upper plane $z=h$ using transfer matrix and taking into account PEC conditions at $z=0$:
\e\label{eq:a9}
E_x(h)=M_{12}H_y(0),\,
H_y(h)=M_{22}H_y(0).\f
Since the area $z>h$ is a semi-infinite material with the relative permittivity $\E_s$ (see Fig.~1), The tangent electric and magnetic field components at $z=h$ are connected via the transverse wave impedance
\e\label{eq:a10}
Z_0=\eta{\sqrt{k^2\E_s-k_x^2}}/{k\E_s}
=i\eta{\sqrt{k_x^2-k^2\E_s}}/{k\E_s}\f
taking into account that we consider waves exponentially decaying from the surface of metamaterial, i.e. $k_x^2>k^2\E_s$.
Substituting into Eq.~\r{a10} expressions for elements of the transfer matrix and transverse wave impedances $Z$ and $Z_0$ we obtain dispersion equation for the TM-modes, propagating along the $x$-axis:
\e\label{eq:a11}
{k_z\tan{(k_zh)}}/{\E_{\perp}}={\sqrt{k_x^2-k^2\E_s}}/{\E_s}.\f
Note, that in our case $\E_{\perp}=\E_h$ due to a small thickness of CNTs and high azimuthal surface conductivity of CNTs (a wave weakly interacts with CNTs if the electric field vector is perpendicular to nanotubes).

\end{document}